# Enhanced Spatially Interleaved Techniques for Multi-View Distributed Video Coding

Nantheera Anantrasirichai and Dimitris Agrafiotis

*Abstract*— **This paper presents a multi-view distributed video coding framework for independent camera encoding and centralized decoding. Spatio-temporal-view concealment methods are developed that exploit the interleaved nature of the employed hybrid KEY/Wyner-Ziv frames for block-wise generation of the side information (SI). We study a number of view concealment methods and develop a joint approach that exploits all available correlation for forming the side information. We apply a diversity technique for fusing multiple such predictions thereby achieving more reliable results. We additionally introduce systems enhancements for further improving the rate distortion performance through selective feedback, inter-view bitplane projection and frame subtraction. Results show a significant improvement in performance relative to H.264 intra coding of up to 25% reduction in bitrate or equivalently 2.5 dB increase in PSNR.**

## I. Introduction

Distributed video coding (DVC) allows shifting of the complexity from the encoder to the decoder making it a particularly attractive approach for low power systems with multiple remotely located encoders, such as multi-camera wireless video surveillance and multimedia sensor networks. DVC stems from information theory results developed in [1] by Slepian and Wolf, and extended in [2] by Wyner and Ziv, on source coding with side information at the decoder. In the multi-view scenario discussed herein, where several cameras capture the same scene from different angles, DVC additionally offers the possibility of exploiting spatial/view correlations among cameras without the need of inter-camera communication, since the prediction (side information) is formed at the decoder side.

A common DVC scenario involves splitting of the video frames into two categories, KEY frames and Wyner-Ziv (WZ) frames [3]-[5]. KEY frames are conventionally intra coded, while WZ frame coding can operate either in the pixel domain where the WZ data are, after quantization, directly encoded in a bit-plane by bit-plane fashion (PD-DVC), or in the transform domain (TD-DVC) whereby a transform is applied prior to quantization in order to remove spatial redundancy. At the decoder, the decoded KEY frames are used to create an estimate of the WZ frames, the side information. The received parity bits are then employed to correct the errors occurring in this noisy version of the WZ data.

For our multi-view DVC codec, we adapt our previously proposed hybrid KEY/WZ frame approach, referred to as SPI-DVC [6]-[8], whereby spatial interleaving (SPI) of KEY and WZ blocks takes place using a chessboard pattern similar to the flexible macroblock ordering (FMO) dispersed pattern of H.264[9]. This then allows us to treat the process of generating the side information (SI) as a block-based error concealment task. The results indicate performance improvements relative to frame based schemes [10]. They also show that the proposed SPI-DVC codec can efficiently handle sequences with non-uniform motion trajectories and occluded objects, both of which represent some of the most commonly encountered difficulties with multi-view systems. In order to avoid any performance loss relative to full frame KEY coding, due to the spatial block interleaving, we additionally employ temporal interleaving (TI) whereby two KEY groups of two consecutive frames are interleaved prior to intra coding. For WZ groups, we use a Gray code which has been shown to improve system performance over a natural binary code [11].

At the central decoder, encoded data streams from all cameras are decoded together using block-based concealment to "conceal" (predict) the missing WZ blocks using information available from the previously received 4-neighbouring KEY blocks. Generally one of the difficulties of multi-view DVC is defining an SI fusion method, as several SI can be generated using temporally (intra-view) and spatially (inter-view) adjacent frames; however, no original frame exists to enable comparison. In this paper, we introduce four joint techniques for the hybrid KEY/WZ approach. The first three techniques fuse the results of temporal and view concealment using i) error measurement; ii) motion/disparity vectors; and iii) intensity and depth consistency. The last joint technique refines the results of view concealment to generate the final SI using temporal concealment. We subsequently introduce a simple and efficient reconstruction method inspired by a diversity technique used in wireless communications. In the DVC scenario this process translates into a multi-hypothesis SI framework. Multiple SI data generate a more precise log-likelihood ratio thereby reducing the number of bits requested by the turbo decoder for correcting the SI [12]. Moreover, the multi-hypothesis approach can reduce the effect of erroneous decisions made during the fusion process. The proposed multi-view DVC codec (referred to as MV SPI-DVC) is illustrated in Fig. 1.

The authors are with the Department of Electrical and Electronic Engineering, University of Bristol, Bristol, BS8 1UB, UK (e-mail: N.Anantrasirichai@bristol.ac.uk).

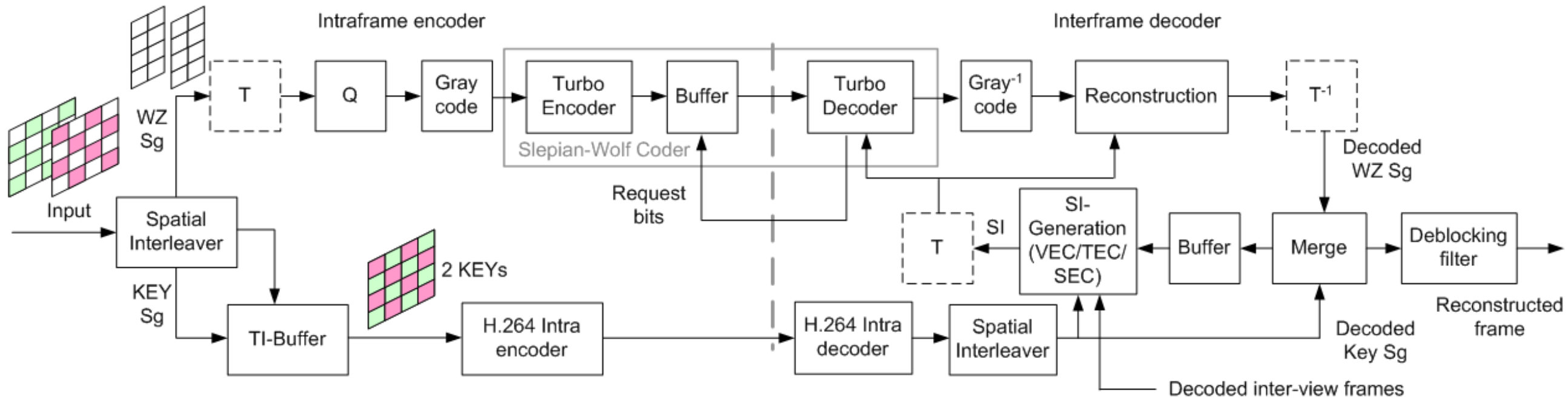

**Fig. 1 The proposed MV-SPI-DVC codec.**

In this paper we also propose and investigate three methods to further enhance the performance of our codec: i) selective feedback for reducing the number of requested parity bits; ii) static area marking for encoding only those blocks that lie in moving regions of the picture; and iii) bitplane projection for exploiting the correlation of associated biplanes among views. One major drawback of many DVC methods compared to hybrid coding approaches like H.264, is that (parity) bits are spent (requested) equally for all regions (blocks) of the frame irrespective of the quality of the local SI (prediction). With the selective feedback approach that we propose the current frame is divided into several regions, the coding order of which is ranked according to the quality of the local SI. The method relies either on reliability maps being sent from the decoder to the encoder or on a prior agreement between encoder and decoder which enables selective parity bit transmissions for those regions that are associated with poor SI. Effectively the same objective is served by the static area marking technique, which, however, requires a slight modification of the encoder for detecting static areas. Bitplane projection on the other hand affects only the complexity of the decoder, through the introduction of an additional depth estimation process. Although the proposed methods increase slightly the overall complexity of the system, they do offer performance improvements that make them worth considering.

The rest of the paper is organized as follows: Section II discusses prior related work. Section III describes the proposed codec, with the GOP structure discussed in subsection III.A, the encoder in sub section III.B and the decoder in sub section III.C. Section IV presents the proposed enhancements. Finally section V gives experimental results, before concluding the paper.

## II. RELATED WORK

Early work on distributed video coding focused on single-view video coding systems. Work on multi-view DVC (MV DVC), although still at an early stage, has already led to the formation of a number of different coding frameworks. Useful reviews on both mono-view and multi-view DVC can be found in [3], [13] and [14], all of which cover many of the existing methods. In the early work of [15] DVC was adapted for use with large camera arrays through the introduction of two types of cameras; conventional cameras producing standard, hybrid coded (e.g. H.264 intra) video and pure WZ cameras producing parity bits that correct the side information generated from the adjacent conventionally coded views at the decoder. In [18] and [19] two purely intra (KEY) cameras are combined with a WZ camera that is generally positioned in-between, and which produces both KEY (intra) and WZ frames in an alternate fashion. In [16] the authors introduce a structure consisting only of N such WZ cameras producing both KEY (intra coded) and WZ frames, with key frames from each camera alternating both in time and space. The same structure is also used in [17]. In this case (as in the case of the single WZ camera of [18] and [19]) the side information generation process at the decoder can use both temporally and spatially adjacent frames for predicting each WZ frame, through the use of temporal motion interpolation (TMI) and spatial view interpolation (SVI) respectively. Existing SVI methods include that of [16] that uses an affine transformation for generating the SI and that of [18] which relies on homography for doing that. More complicated techniques have been introduced that make use of depth information, whereby the current view is interpolated using view synthesis techniques [20], geometric-based rendering [15] or projection from 3D point clouds [19].

Exclusive use of TMI or SVI for the generation of the side information is problematic. TMI can fail when there is high motion between frames or when the motion is not purely translational. SVI on the other hand will fail in the presence of occlusions or other problematic features such as reflections leading to false depth information. It is thus necessary to employ a fusion approach for generating the SI that will assess the reliability of the pixels estimated by each of the two methods. Typically such fusion methods employ a binary reliability mask for indicating which pixels should be generated with TMI and which with SVI. Since the original version of the WZ frame is not available, the previous, future or spatially adjacent KEY frames are used for calculating the fusion mask [16][17][19], often involving view projection. An alternative technique [18] uses the magnitude of the motion vectors resulting from TMI as a fusion criterion, based on the assumption that temporal estimation generally performs poorly in regions of the frames where motion is significant. Weighed fusion through reliability masks is proposed in [17], where it becomes clear that the generation of the binary fusion mask influences the SI quality thereby driving the system performance. Despite the above advances, the rate distortion performance of most multi-view DVC systems is still far from that achieved with H.264 simulcast.

## III. THE PROPOSED CODEC

The proposed framework of spatially interleaved DVC for multi-view distributed video coding (MV SPI-DVC) is described in this section. First the group-of-picture (GOP) structure used in our codec is described. Subsequently the operation of the encoder is discussed, including the interleaving step, coding considerations for the KEY blocks and finally transformation and quantization of the WZ data. This is then followed by a description of the decoding process which includes the side information generation, frame reconstruction and estimation of the correlation noise.

### *A. GOP Structure*

The proposed MV SPI-DVC uses an IB coding structure. The first frame of each GOP is coded using Intra-frame coding (I-frame) and acts as a refresh frame that stops any "error-propagation" due to uncorrected errors still present in the WZ data after channel decoding. The rest of the frames in the GOP are coded using a combination of KEY and WZ coding, the latter involving interpolation at the decoder for generating the SI (B-frame).

To maximize the prediction efficiency and minimize error accumulation, the distance between the B-frames and the I-frames is globally minimized by shifting the positions of the GOPs in the even views. That is, intra-frame coding in the odd views is applied to the first frame and the frames ($n{\cdot}L_{GOP}$+1), while intra-frame coding in the even views is applied to the first frame and the frames ($n{\cdot}L_{GOP}+\lceil(L_{GOP}+1)/2\rceil$), where $n \in \{0,1,\ldots\}$ and $L_{GOP}$ is the GOP length. The B-frame with the shortest distance to I-frames is decoded first in order to achieve high-quality decoded images, as the shorter distance between the current frame and the reference frames presents higher correlation between them. However, if the surplus delay is of concern, then B frames can be sequentially decoded from the first B frame of the GOP to the last frame. Clearly, this structure is more practical if the GOP length increases. Sample GOPs of 2 and 3 frames are illustrated in Fig. 2, in which the white blocks and the chessboard blocks represent the I-frames and the B-frames, respectively.

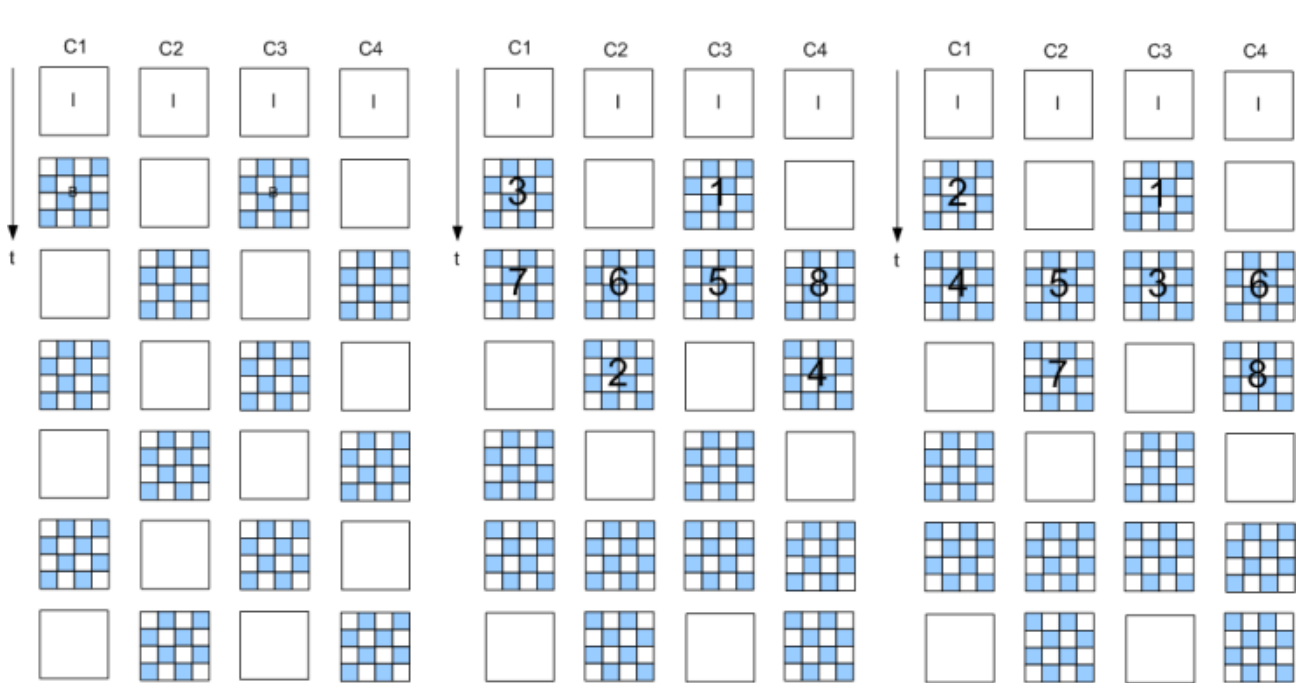

(a) (b) (c)

**Fig. 2 GOP structure for four views for a GOP length of a) 2, b) 3 with minimized IB distance and c) with minimized delay. The numbers overlaid on B-frames indicate coding order.**

### *B. Encoder operation*

The cameras used in a multi-view DVC system do not generally communicate with one another. Each encoder therefore functions as a single view encoder. For general details and parameters associated with the operation of the proposed encoder we refer readers to [7][8]. Below we only describe some KEY-coding aspects of the encoder.

#### *1) Spatio-Temporal Interleaving and KEY group coding*

The first step in the proposed framework involves splitting the current input frame into KEY and WZ groups, in a similar fashion to the dispersed type of FMO specified in H.264 [9]. Generally the block size (i.e. the interleaving step) is fixed for the entire input sequence and can range, for example, from 16x16 to 4x4 pixels. The choice of block size can affect the coding performance of both KEY and WZ data and thus is the first encoder design aspect that needs consideration. Interleaving steps smaller than 4x4 are not of interest as they can interfere with the de-correlating ability of the 4x4 transform that is applied to both types of data in the case of transform domain SPI-DVC. Bigger block sizes can benefit KEY (H.264 Intra) coding as they allow better utilization of context information by the context adaptive entropy coding module (CAVLC or CABAC), as well as better de-correlation of the DC coefficients than the 4x4 blocks. However, regarding WZ coding (or rather decoding) and given the concealment oriented approach of the proposed system, we expect smaller block sizes to benefit the performance more, as better prediction (concealment) of the SI should be achieved especially in areas of high motion (or high texture variance when spatial concealment is used). The interleaving could also be beneficial to the performance of the turbo decoder as suggested in [22] by preventing the formation of continuous pixel/bit runs from low correlation regions that can represent burst errors to the turbo decoder if the SI generation fails.

After the interleaving process, the KEY groups are, normally, horizontally shifted to make a new frame of the same height but half the width of the original which is then encoded with H.264 in intra mode. The H.264 de-blocking filter is not applied to the KEY groups, as the interleaving process creates artificial neighbors. Filtering across such block edges creates perceptual artefacts and does not benefit the prediction process at the decoder (where de-interleaving will take place). Note that a post-processing H.264-like de-blocking filter can be used at the decoder for smoothing out blocking artefacts of both KEY and WZ reconstructed blocks.

If the GOP length is more than 2, the KEY and WZ groups alternate relative to the previous frame so as to avoid creating potentially annoying regions of different subjective quality. The KEY groups of two successive frames are combined (as shown in Fig. 3) to avoid any significant performance loss relative to full frame intra KEY coding, especially when the spatial prediction modes of H.264 are employed. Fig. 4 shows the improvement in KEY coding performance when the KEY groups of two successive frames are combined before applying H.264 Intra coding. The performance loss compared to full frame intra coding has now become insignificant.

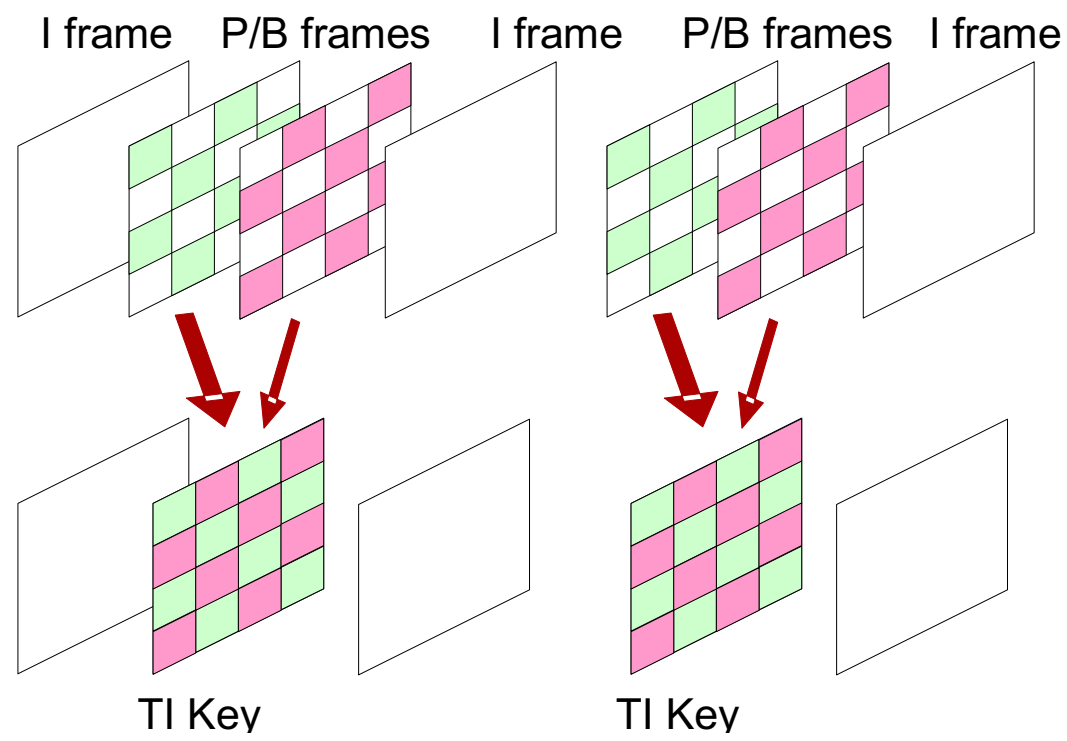

**Fig. 3 Temporal interleaving of KEY groups for a GOP of 3.**

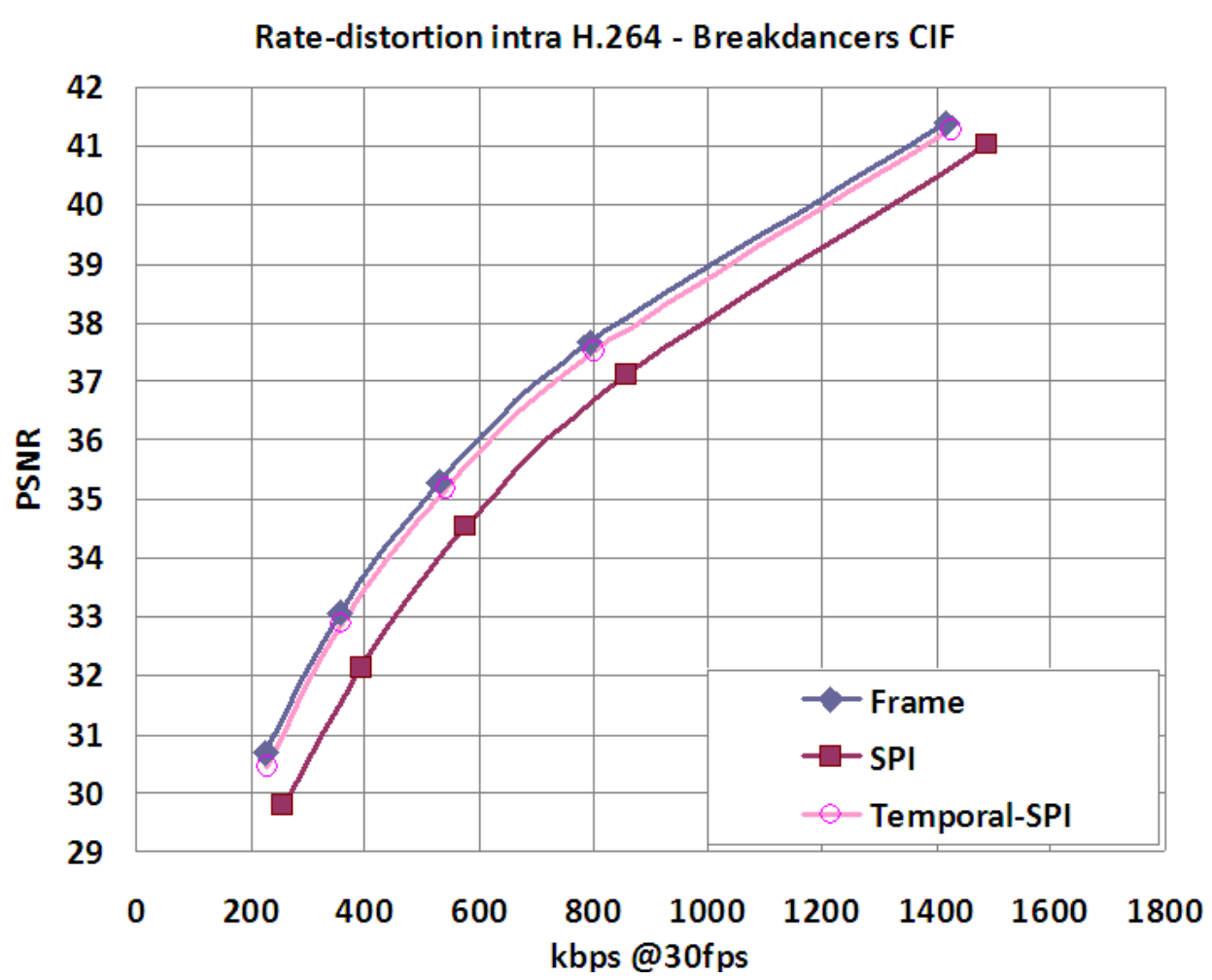

**Fig. 4 Full frame and KEY group H.264 intra coding performance with and without temporal interleaving of KEY groups.**

*2) Gray Code and Wyner-Ziv Group Coding*

After transformation and quantization, the quantized symbols are converted into binary data. Subsequently, the binary codes are converted to Gray codes by XORing the binary values with their logical-shift-right values. At the decoder, the decoded Gray codes are converted back to binary values in a similar fashion. The Gray code conversion clearly adds to the complexity of the system, but it significantly improves the codec's performance as it has a Hamming distance of one which provides for higher error resilience than the natural binary code, and offers a higher accuracy of bit probability estimation. The WZ data are fed to a Turbo coder in a bitplane-by-bitplane fashion, and then only the parity bits are transmitted depending upon requests from the decoder.

## C. *The multi-view SPI Decoder*

The decoder works in a centralized manner. The received data from all cameras are jointly decoded. The KEY data are used for creating the side information. This side information is seen as the systematic part of the channel encoder's output as received at the decoder, i.e. the SI is seen as a noisy version of the coded WZ frame. The received parity bits are used to correct the errors in this noisy version of the WZ data.

The side information generation process is an important part in the design of a DVC system as it directly relates to the system's rate distortion performance. The better the side information, the fewer the parity bits required to decode the WZ data correctly. Generally in single-view video coding, the decoder employs motion estimation and compensation which use previously decoded past and future frames for creating the SI depending on the delay constraints of the application. In the multi-view scenario, the correlation across multiple view geometry is also exploited to improve SI quality. One way the latter can be achieved is by improving prediction in the presence of occlusions due to motion, as such occluded areas can potentially be visible in other views.

In the proposed hybrid KEY/WZ framework, the generation of the SI is equivalent to an error concealment process for missing blocks (WZ blocks) in the presence of all their 4-neighbours (KEY blocks). We employ view (disparity or geometric), temporal and spatial error concealment methods (VEC/TEC/SEC) the application of which is controlled by a mode selection algorithm. A detailed description of the TEC and SEC methods used can be found in [22]. Below we briefly describe how they are applied in the context of B frame SI generation for each WZ block. We discuss VEC and joint EC methods in more detail.

*1) SI using TEC*

The employed TEC method uses the external boundary matching error (EBME) of a WZ block -defined as the sum of absolute differences (SAD) between the multiple pixel boundary of blocks adjacent to the missing one (WZ) in the current frame and the same boundary of blocks adjacent to the replacement block in the reference frame (see Fig. 5). Spatially adjacent MVs are generated through bidirectional motion estimation (ME) for all blocks of the KEY group ($MV_{Kn}$) prior to the concealment process. The search range of this motion estimation is adjusted according to the distance between the current and the reference frame. 1/4-pel motion vector resolution is used for better ME performance. The initial estimation of a WZ motion vector is further refined through overlapped block motion compensation (OBMC).

*2) SI using VEC*

For view error concealment we investigate three approaches based on disparity, homography and depth estimation, the performance of which we compare. The simplest approach for exploiting the inter-view correlation is disparity estimation and compensation. The concealment process in this case is similar to the temporal concealment method described above, with the only difference being the use of adjacent views as reference frames. The minimum SAD is used for determining the best replacement block resulting from the disparity search. This method is straightforward to implement and only requires an increase in buffer size and computation time.

The geometric relationship between multiple cameras can also be utilized to estimate the missing blocks. A popular approach with frame-based multi-view DVC coding schemes in the literature is the use of homography, in which the projection matrices of cameras are calculated once, at the beginning of the sequence, and are subsequently used for the rest of the sequence [16][18]. In our framework SI generation through homography can be applied more efficiently as it is done at the block level. The neighboring KEY blocks provide local information for calculating an individual transformation matrix for each WZ block.

A more complex approach to exploiting adjacent views for SI generation makes use of depth information for interpolating the WZ camera frames. Using camera parameters, each pixel in the intra camera frame is projected into 3D space, and then, given the WZ camera parameters, this 3D representation is projected as a 2D image onto the WZ camera [19]. For our test simulation, we assume that the camera parameters are known. We employ a dynamic programming approach for depth estimation as it offers better results compared to a traditional pixel-based approach [24]. Given that at least two views are required for depth estimation we resort to TEC for those frames that have zero or one adjacent intra frame available.

We tested the aforementioned algorithms using the Breakdancers and Ballet sequences. Fig. 6 illustrates the results, including results obtained with TEC for comparison. The left plots display the quality of the side information (mean square error - MSE), whilst the right plots show the quality of SI only for those areas where TEC doesn't perform very well (a pixel luminance error of more than 10). The plots on the left show that the Breakdancers sequence contains higher inter-view correlation than temporal correlation as the results of TEC have a higher MSE compared to those of VEC (disparity and depth). The opposite is true for the Ballet sequence. The plots on the right however suggest, that VEC can be beneficial for decoding both sequences, as VEC can improve the quality of those regions in the SI that TEC doesn't predict adequately, and particularly so at high bitrates. Homography-based estimation gives the worst results as the test sequences do not satisfy the homography projection conditions (identical cameras with same centre of projection and no translation). The disparity approach seems to provide the best VEC results for the Ballet sequence, and nearly so for the Breakdancers sequence, for which depth estimation performs best. However given that in reality depth maps would have to be estimated from compressed views received at the decoder, leading to both inaccurate estimates and increased complexity, we prefer to use disparity estimation for performing VEC.

*3) SI using Joint TEC and VEC*

Following the above results, we now focus on methods of combining/fusing TEC and VEC. Because our framework is based around the hybrid KEY/WZ frame, it lends itself easily to an efficient block based fusion method. We study a number of ways for deciding between TEC and VEC in order to create the best possible SI. The presence of the KEY data in the current frame is exploited for identifying regions of occlusion and prediction failure in general.

The simplest approach to selecting between TEC and VEC is through the matching error used in the prediction and concealment process (whole block SAD for KEY blocks and EBME for WZ blocks). The method with the smallest matching error is chosen for predicting or concealing a specific KEY or WZ block respectively. A high value in the matching error implies a mismatch in the motion/disparity estimation process which often occurs in occluded areas.

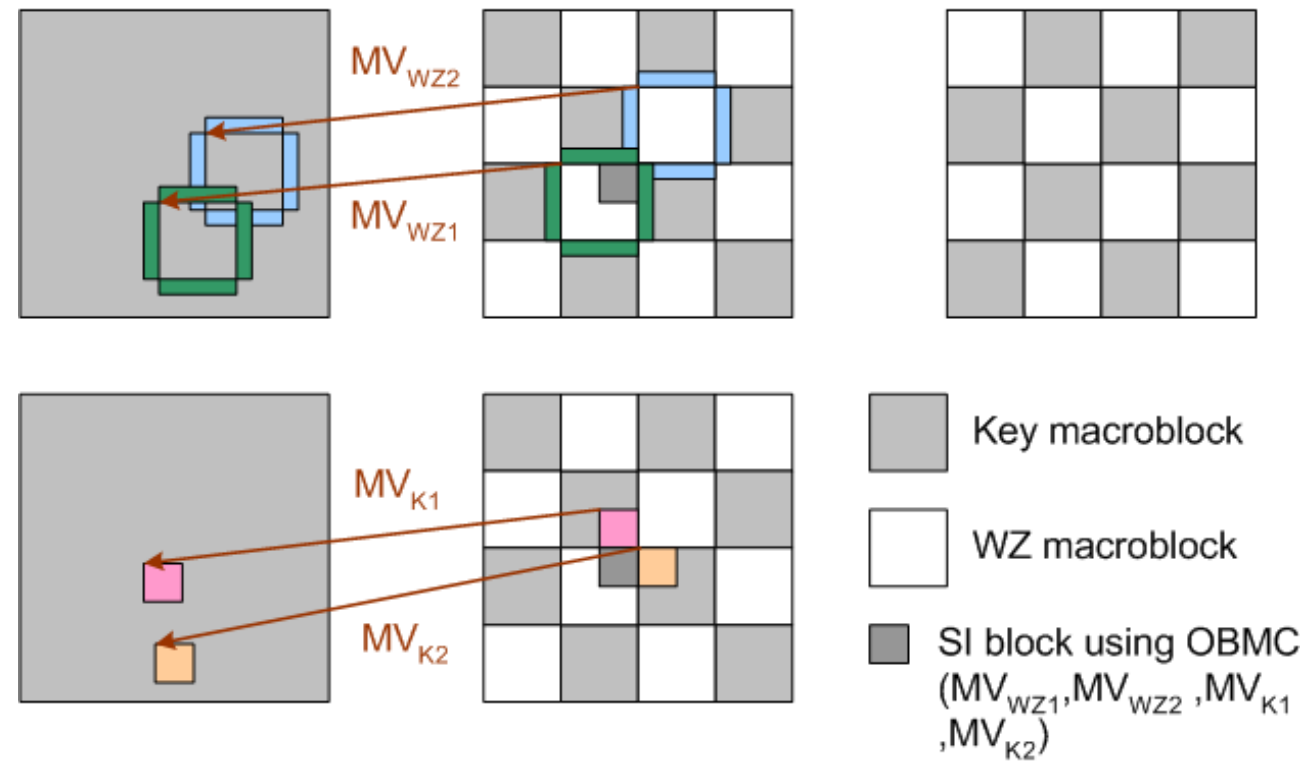

**Fig. 5 Example of the TEC process.**

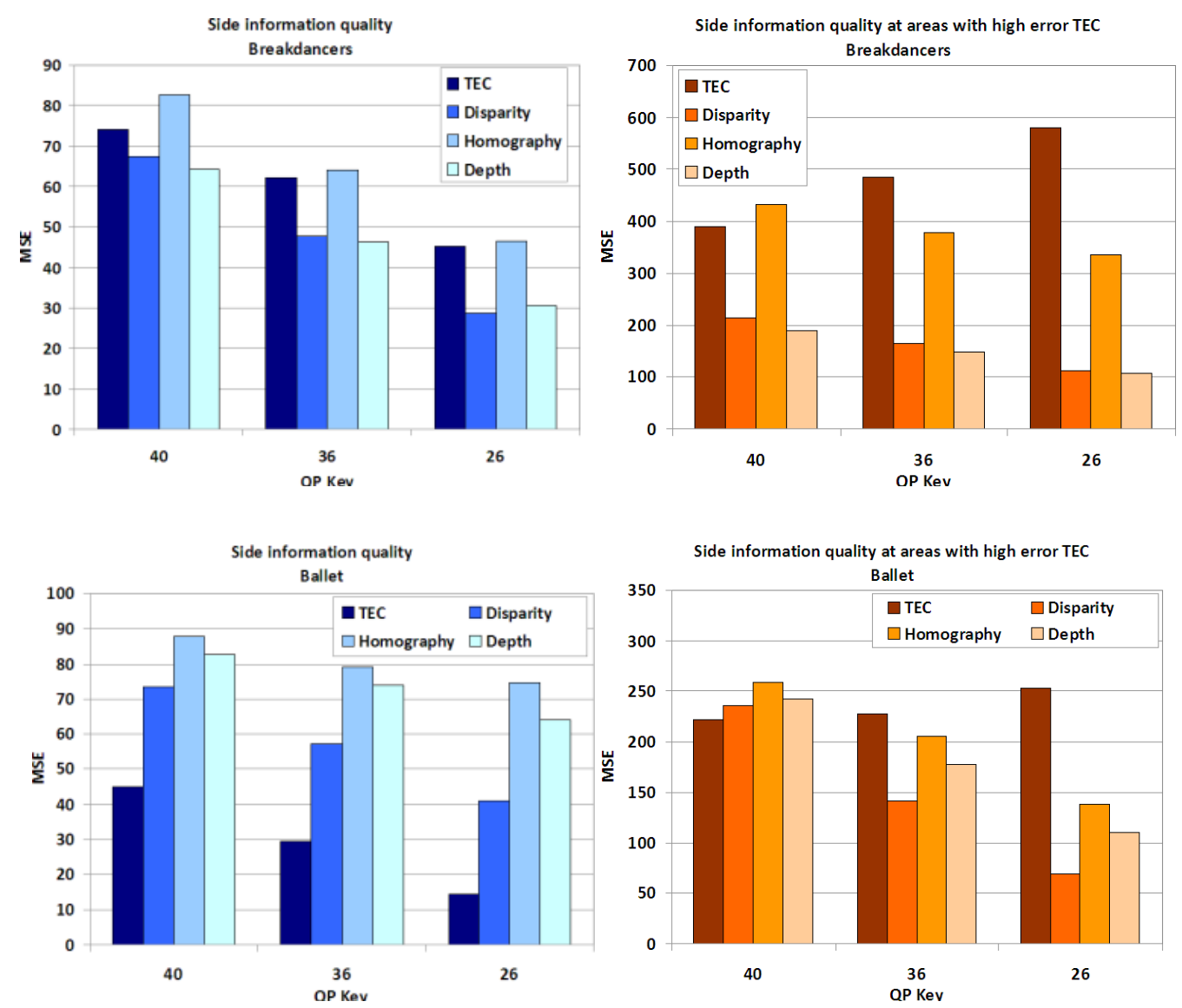

**Fig. 6 Comparison of achieved SI quality using VEC.**

A second possible fusion method relies on the uniformity of neighboring motion and disparity vectors for detecting occlusions. Vectors varying significantly from those estimated for neighboring blocks are considered outliers and are taken to suggest occlusions. The distance of a block vector from the median value of a group of neighboring vectors is used for detecting such blocks. For each KEY block, the 4-neighboring KEY blocks and 2 neighboring WZ blocks, together with the examined block vector are used for the median calculation. A similar neighborhood is used for each WZ block. We set a distance threshold for the estimated motion vector which, when exceeded, sets off a decision algorithm. This compares the distance of the estimated block motion and disparity vectors with their respective neighborhood median and selects the prediction method that results in the smaller distance.

The hybrid KEY/WZ framework allows further enhancement of the fusion process if depth information is available. Using luminance and depth consistency constraints, the projected view is compared to the KEY group of the current view. Only the WZ blocks for which the neighboring KEY blocks conform to these constraints are concealed using VEC. We set two thresholds for the luminance and depth constraints which here are 15 and 0.04 respectively.

A fourth and final method for combining VEC with TEC involves the use of the side information generated by the former as input to the latter. This initial VEC-produced SI is employed together with the external boundary of each WZ block in the motion estimation (temporal concealment) process of each WZ block. The external boundary matching error measure of [24] is now modified as follows:

$$\begin{aligned}\mathrm{EBMC}^m = & \sum_{i=1}^{M}\sum_{x=x_0}^{x_0+N-1}\left[\left|F_{x,y_0-i} - F^r_{x+u_x,y_0-i+u_y}\right| + \left|F_{x,y_0+N-1+i} - F^r_{x+u_x,y_0+N-1+i+u_y}\right|\right] \\ & + \sum_{i=1}^{M}\sum_{y=y_0}^{y_0+N-1}\left[\left|F_{x_0-i,y} - F^r_{x_0-i+u_x,y+u_y}\right| + \left|F_{x_0+N-1+i,y} - F^r_{x_0+N-1+i+u_x,y+u_y}\right|\right] \\ & + w^m \cdot \sum_{i=0}^{N-1}\sum_{j=0}^{N-1}\left[\left|F^{SI}_{x_0+i,y_0+j} - F^r_{x_0+i,y_0+j}\right|\right]\end{aligned} \quad (1)$$

where ( $x_0, y_0$ ) is the coordinate of the top left pixel of the WZ block, ( $u_x, u_y$ ) is the candidate vector under consideration, $N$ is the block size, $M$ is the width of the boundary, and $F_{x,y}$, $F^r_{x,y}$ and $F^{SI}_{x,y}$ are the pixels of the current, the reference and the SI frame, respectively. $w^m$ is an adaptive non-negative weight varied upon the local SAD generated from the VEC method. If the initial SI block is associated with a large prediction error at the neighboring KEY blocks, then $w^m$ is small, and vice versa. Herein we use the function $1-\left(\alpha \cdot e^2 / 1+\alpha \cdot e^2\right)$, where $e$ is the SAD value of the initial SI and $\alpha = 1/(n \cdot th)^2$, where $n$ and $th$ represent the number of pixels in the external boundary and error threshold respectively (a value of 5 is used in this paper). Two flavors of this refinement approach are examined. One where TEC is used by default for the final concealment and one where the same decision criterion as in the first SAD-based method is used to decide between VEC and TEC.

Fig. 7 shows the quality (PSNR) of the side information when generated using one of the four combination/fusion techniques described above. The matching error based fusion (SAD) performs best for both tested sequences. Fusion based on depth information and projection (fusion Depth) performs worse as does fusion based on motion/disparity vector uniformity. Both however improve the quality of the SI compared to using one EC method alone (TEC for Breakdancers, VEC for Ballet). The use of the VEC-produced SI does improve the TEC performance in the refinement technique (refine SI) but it does not offer any further benefits when combined with a matching error based decision on the preferred type of EC (fusion refine SI), in which case it performs similarly to the matching error alone based approach. It is clear that a joint VEC/TEC approach benefits the quality of the SI significantly especially at higher bitrates where finer quantization can exaggerate any prediction mismatch.

*4) SI using SEC*

In the case of hybrid coding a larger search window can lead to better motion estimation and thus performance. In the DVC case however this is not always the case as the original (WZ block) is not known. Hence, spatial error concealment (SEC) is necessary in cases where the motion information is not reliable. We employ the method of [25], whereby bordering KEY pixels are used for concealing the missing pixels of a WZ block through bilinear interpolation (BI) or directional interpolation (DI) along detected edges. The type of interpolation used depends on the outcome of a decision algorithm that uses the directional entropy of neighboring edges for choosing between the two interpolation approaches. For more information on the SEC module the reader is referred to [25].

*5) SI concealment mode selection*

The mode selection algorithm examines the suitability of the joint TEC/VEC method for concealing each WZ block, by evaluating the levels of motion/disparity compensated activity (MDA) and spatial activity (SA) in the neighborhood of that block and switching to spatial concealment accordingly. Motion/disparity compensated temporal activity is measured as the mean squared error between the KEY blocks surrounding the missing one in the current frame and those surrounding the replacement block in the reference frame. Spatial activity is measured as the variance of the surrounding KEY blocks in the current frame. More formally:

$$SA = E\,[\,(x-\mu)^2\,] \text{ and } MDA = E\,[\,(x-x^*)^2\,] \quad (2)$$

where $x$ are the pixels in the neighborhood of the missing block and $x^*$ are the pixels in the neighborhood of the replacement block in the reference frame. SEC is employed if the spatial activity is smaller than the temporal/disparity activity and the latter is above a threshold (3 in this work).

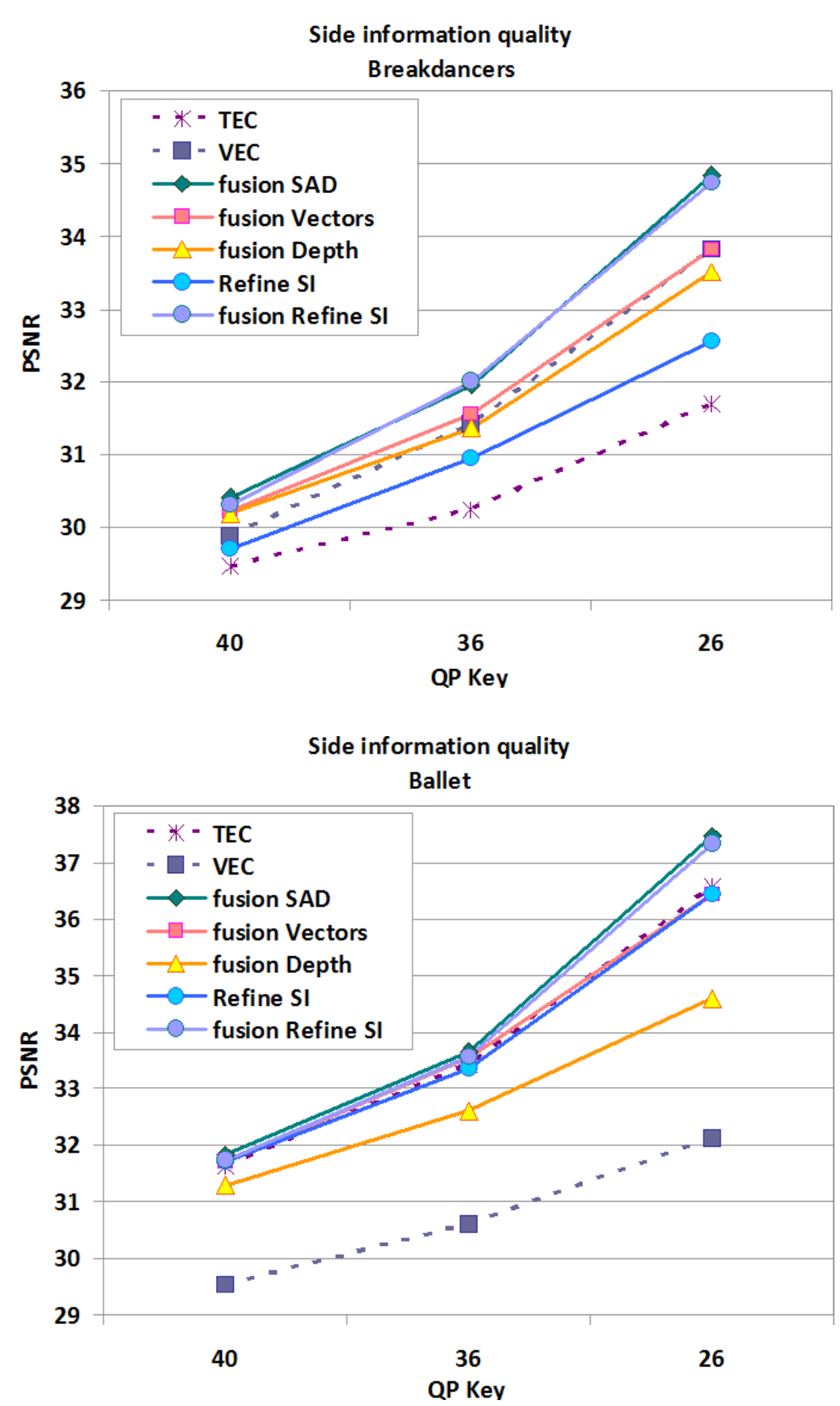

**Fig. 7 SI PSNR for different VEC/TEC fusion methods.**

*6) Multi-hypothesis decoding and reconstruction*

To increase the accuracy of the statistics used in the turbo decoding, we use a diversity technique, similar to those employed in wireless communication systems. Diversity techniques exploit multiple received signals (e.g. through the use of multiple antennas) to combat problems caused by channel fading. In the DVC scenario, this process translates into a multi-hypothesis SI framework, wherein multiple SI data available at the decoder are used to determine the conditional probability function of the source. A more precise log-likelihood ratio is consequently computed thereby reducing the number of requested bits for correcting the SI.

Multiple SI is generated and defined as $SI_i$, $i \in \{1,..,N\}$, with increasing $i$ denoting the ranking of the SI, from best to worst, according to the MSE of the predicted KEY group associated with each SI group. Previously reconstructed WZ data based on available decoded bit planes are additionally used in the LLR computation expressed as:

$$LLR^l = \log \frac{\Pr\left[X^l = 0 \middle| SI_1, SI_2, ..., SI_N, Y^{l-1}\right]}{\Pr\left[X^l = 1 \middle| SI_1, SI_2, ..., SI_N, Y^{l-1}\right]} \quad (2)$$

where $X^l$ is a bit value at bit-plane $l$ and $Y^{l-1}$ is a reconstructed WZ data $Y$ using decoded bit-planes ($l$-1). Note that the turbo decoder is sensitive to extreme LLR values, and we therefore average as opposed to summing the value of the multiple LLRs. The decoded bit-planes are converted to quantisation symbol $\widetilde{Q}$. The pixel $Y$ is then reconstructed by firstly using the best SI based on its quantisation bins as follows:

$$Y_1 = \begin{cases} \Delta Q \cdot (\widetilde{Q}+1) & Q_{SI_1} < \widetilde{Q} \\ SI_1 & Q_{SI_1} = \widetilde{Q} \\ \Delta Q \cdot \widetilde{Q} & Q_{SI_1} > \widetilde{Q} \end{cases} \quad (3)$$

where $\Delta Q$ is a quantisation step size and $Q_{SI}$ is the quantisation bin of the SI. Next, the pixels of $SI_2$ with $Q_{SI_2} = \widetilde{Q}$ are used to replace the clipped pixels (where $Q_{SI_1} \neq \widetilde{Q}$). The other SI is sequentially applied to the clipped pixels as follows:

$$Y_p = \begin{cases} SI_p & Q_{SI_p} = \widetilde{Q} \text{ and } Q_{SI_{p-1}} \neq \widetilde{Q} \\ Y_{p-1} & \text{otherwise} \end{cases} \quad (4)$$

where $Y_p$ is the pixel value after the $p^{\text{th}}$ SI has been employed, $p \in \{2,..,N\}$. In this paper we use three such SI data; $SI_1$ is the best estimate resulting from the use of TEC/VEC/SEC and mode selection; $SI_2$ is the intra-view SI (generated from TEC alone); and $SI_3$ is the inter-view SI (generated from VEC alone). Having multiple SI data can compensate for potential errors in the fusion process (Fig. 8).

*7) Correlation Noise Estimation*

Aside from the SI generation, the other big factor affecting the performance of DVC coding is the estimation of the virtual dependency channel, i.e. the correlation noise that arises from the difference of the generated side information and the original WZ data. It is common to assume that this noise follows a Laplacian distribution with a specific variance that changes from frame to frame (and even between regions of the frame). Some authors [10] assume knowledge of the channel at the decoder which allows them to specify a value for the Laplacian distribution parameter $\alpha$, where $\alpha$ is equal to $\sqrt{2}/\sigma$, with $\sigma$ being the standard deviation of the noise. This of course is not a realistic scenario as it requires knowledge of the original WZ pixels at the decoder. A more realistic approach is that of estimating the value of alpha from a locally generated distribution that should normally approximate that of the correlation noise. The better this approximation is the better the alpha estimation and as a result the better the operation of the turbo decoder.

We estimate the correlation noise on a block basis using the 4-neighbouring KEY blocks of each WZ block. The noise is modelled as a Laplacian distribution with a specific variance that changes from block to block. The estimation of this variance is performed in one of two ways depending on the outcome of the concealment mode selection algorithm. If the TEC/VEC is selected, we employ the difference between the 4-neighboring KEY blocks and the corresponding motion/disparity compensated blocks in previously decoded frames that are found to provide the best match during the motion/disparity estimation process, in order to estimate the Laplacian distribution parameter-$\alpha$. In other words, having performed motion/disparity estimation for the KEY blocks of the current frame, we take the difference between each KEY pixel and its best match in one of the reference frames. If the proposed concealment algorithm chooses SEC for concealing the current block, we employ the variance of the neighboring KEY blocks for estimating $\alpha$. The fact that SEC is employed when the spatial activity, measured as the variance of the neighboring KEY blocks, is low, can result in very high values of $\alpha$.. To prevent numerical overflows, we limit the maximum value of alpha for WZ blocks concealed with SEC to 0.5.

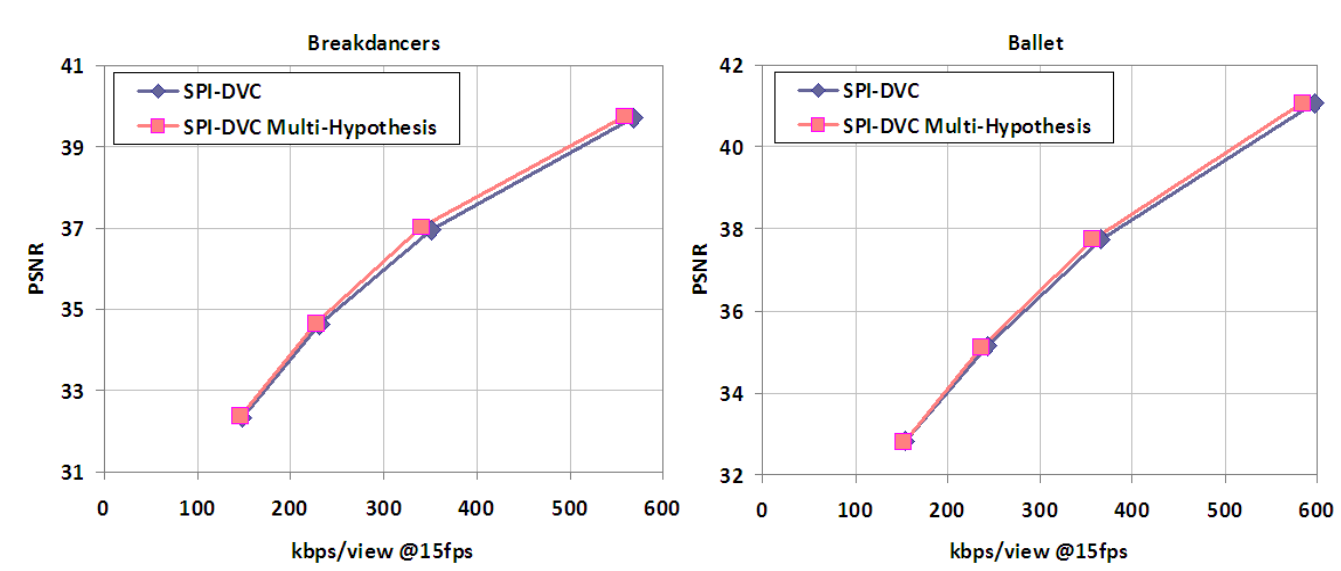

**Fig. 8 Multi-hypothesis decoding / reconstruction performance.**

## IV. ENHANCEMENTS

We propose and investigate three methods for enhancing the performance of the multi-view SPI-DVC codec, namely selective feedback, frame subtraction and bitplane projection.

### A. *Selective feedback*

Generally, at each bit-plane, the decoder requests a set of parity bits from the encoder via a feedback channel whenever the estimated bit error rate (BER) is more than a threshold. In most DVC systems this request for additional parity bits involves the update of all regions of the coded frame irrespective of the quality level of the local SI. As a result unnecessary bits are sent leading to a significant increase of the total bitrate. We propose two methods for introducing some granularity to this feedback process. These two methods are referred to as simple and smart, according to the additional complexity that they introduce to the encoder. They both rely on a prior division of the WZ data at the encoder into several groups which are coded separately. Note that in the transform domain this approach is not applied to the DC coefficients, and for the rest of the frequency bands a prior grouping according to quantization level takes place.

*1) Simple encoder approach*

At the decoder, a feedback buffer is employed for collecting successive clipping maps for a number of consecutive frames which identify pixels with $Q_{SI} \neq \tilde{Q}$ for each decoded bit-plane (12 frames are used herein). These WZ pixels are then segmented into ($BP_{max}$) groups, where $BP_{max}$ is the total number of bit-planes used in WZ coding, with each group signifying the bit-plane where reconstruction mismatch first took place. An additional group is formed for areas where $Q_{SI} = \tilde{Q}$ for all bit-planes. The WZ pixels are then segmented on a 16x16 block basis according to these pixel groups. Parity bit requests are then made at the group level until the desired BER is reached, with priority given to the most significant bit-plane. A tradeoff between update frequency and segmentation accuracy can clearly be made. An example is given in Fig. 9.

The WZ data is quantized into 3 bitplanes. The feedback map is thus divided into 4 groups. The decoder uses the clipping maps to decide which group each 16x16 block belongs to starting from the most significant bitplane. If one or more pixels in the block of the clipping map indicate $Q_{SI} \neq \tilde{Q}$, this block is set to group 1. The blocks that are not yet categorized are re-checked with the clipping map of the second bitplane. This process continues until the completion of group 3. The uncategorized blocks are then set to belong to group 4. The order of the parity bit requests is as follows. Parity bit requests for group 1 are made first and continue until the group's BER is below the set threshold. Parity bits for group 2 are then requested, if the BER of the whole frame is bigger than the desired one and so on. The requests for parity bits are terminated when the BER of the frame reaches the set threshold. In the example of Fig. 9 the desired BER is achieved during the transmission of group 2 parity bits. Therefore, no parity bits for groups 3 and 4 are requested.

*2) Smart encoder approach*

The previous approach relies on the decoder sending update maps to the encoder. To avoid this additional overhead we propose another approach (Fig. 10). Grouping of the WZ pixels into fixed size blocks (e.g. 32x32) first takes place. Parity bit requests are then made for each block in raster scan order. Both the encoder and decoder keep track of the number of parity bit requests made for each block, thus building a block-wise reliability map for each frame with blocks being ranked according to parity bit requests. This ranking is then employed for deciding the coding order of those blocks in the following frame, starting with the region with the largest number of requests. The block BER is calculated and used to decide when to move to the next block in the coding order, and the frame BER to decide when to stop requesting bits for the entirety of the frame. In Fig. 11 we demonstrate examples of the feedback map and the coding order used in the simple and smart encoder approaches, respectively. Both maps show that the middle of the frame, where the moving object is located, is coded before the rest of the frame. Clearly this idea can also be applied to the mono-view scenario.

*B. Bitplane projection*

For this method some cameras are selected to transmit only the KEY data: KEY frames and KEY slice groups. At the decoder the B-frames of these cameras are reconstructed using depth information, KEY data and reconstructed frames from other views. Note that as at least two views are conventionally coded, the decoder has enough information to estimate the depth of the scene. The SI of the WZ data is generated as before, but is corrected using decoded bitplanes projected from neighboring views instead of parity bits from the encoders. The reliability of the projected bitplanes is checked, since the estimated depth can contains errors.

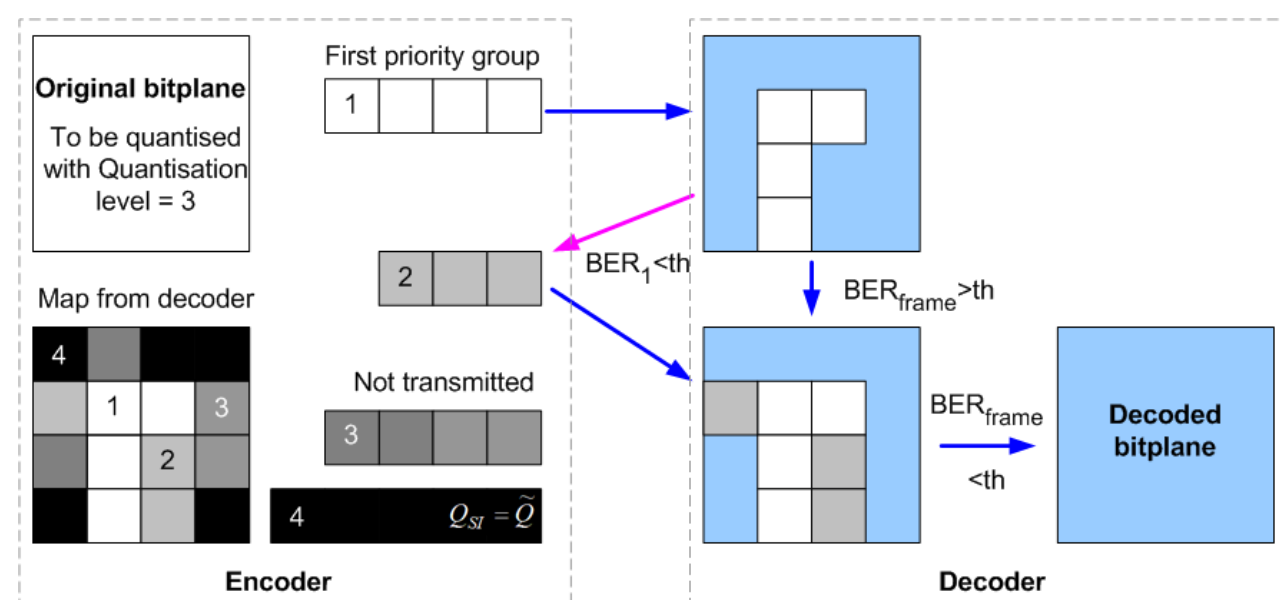

**Fig. 9 Selective feedback- simple encoder: Coding order example.**

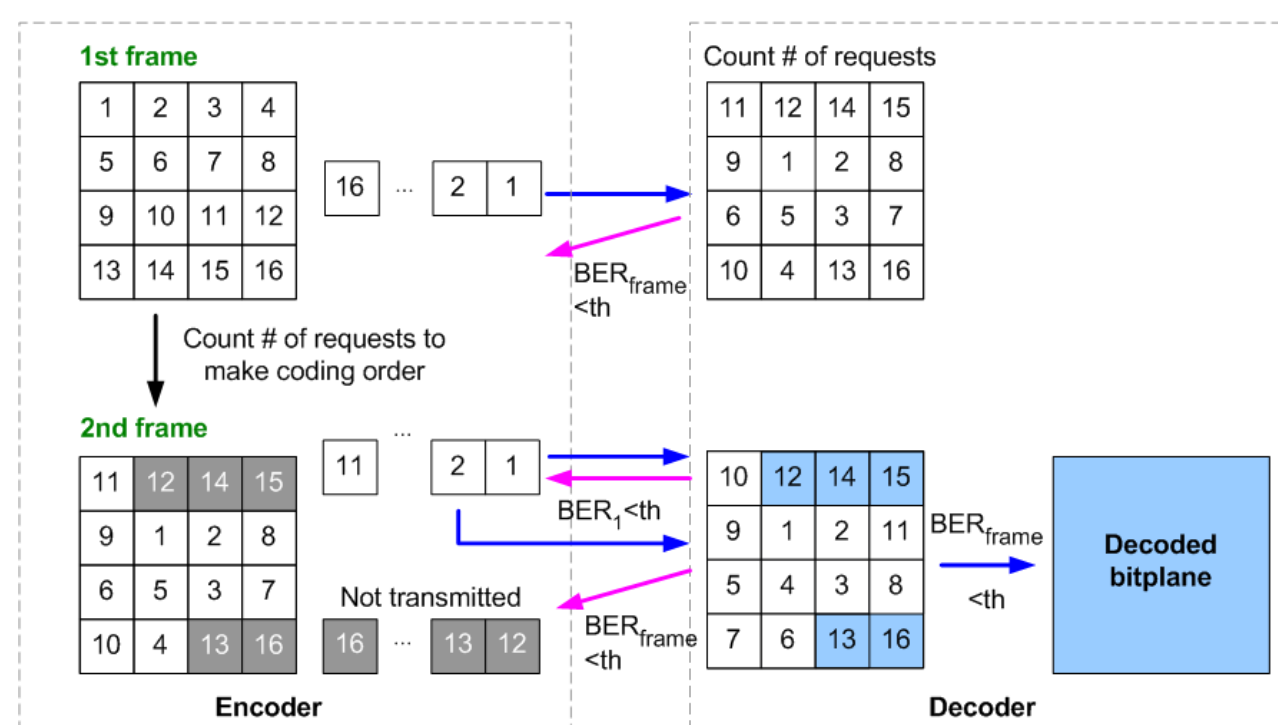

**Fig. 10 Selective feedback- smart encoder: Coding order example**

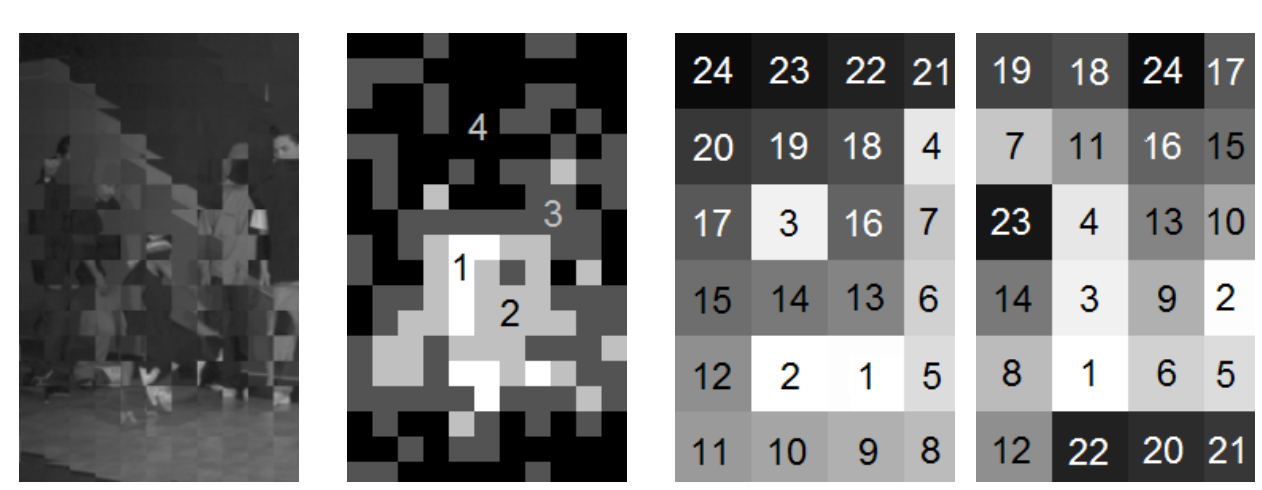

(a) (b) (c)

**Fig. 11 Selective feedback for one frame of Breakdancers: a)WZ data, b) feedback map for simple encoder, c) coding order with smart encoder for two bitplanes**

We propose two approaches for checking reliability; luminance/depth consistency or first BP agreement.For the luminance/depth consistency, two thresholds are set (15 and 4 herein) for luminance and depth respectively. If the difference in luminance and depth between the projected view and the KEY data available in the neighborhood of a particular WZ block are within the thresholds, the projected bitplanes are used to reconstruct the block. A more reliable decision can be made if the first WZ bitplane is decoded. The decoder requests the parity bits of this bitplane, and then uses the reconstructed bitplane to check whether the projected first bitplane is identical. If so, the rest of the projected bitplanes are employed.

### C. Frame subtraction

The lack of any form of prediction and mode selection at the encoder is one of the features that make DVC an attractive approach for low/limited power systems. If a minor increase in complexity is possible then one way of exploiting it is through frame subtraction (prediction with zero motion). Given that most multi-camera systems are installed with fixed camera positions, frame subtraction has the potential of identifying parts of a static background (although perhaps not all as the environment is not controlled). For each macroblock of a B frame consisting of two KEY blocks and two WZ blocks, the sum of absolute differences (SAD) between two successive frames is compared with a predefined threshold. Macroblocks with SAD above the threshold are defined as changing (dynamic) areas and are consequently encoded. A map is created with 1s representing the dynamic macroblocks and 0s the static ones. The dynamic macroblocks of the first frame of two consecutive B frames are split into KEY and WZ blocks. Then, the dynamic KEY blocks are overlaid on the reference I frame to create a new reference frame for the second B (B2) frame. This new reference should provide a better prediction for B2 as it has been updated with the more recent dynamic KEY blocks. After detecting the static blocks of the second B frame (B2), the two sets of dynamic WZ data are combined and coded together via WZ coding. The two maps generated can be transmitted with or without prior encoding (e.g. JBIG). Two sets of KEY data are also combined to create a new KEY frame. The empty blocks, which correspond to the static blocks not being sent, are filled with the mean value of the dynamic blocks. This KEY frame is then encoded with H.264 Intra coding. At the decoder, the blocks marked as dynamic (marked with 1) are reconstructed as explained in previous sections, while the rest of the blocks are replaced by the co-located blocks in the reference frame.

## V. RESULTS AND DISCUSSION

The proposed MV SPI-DVC was evaluated using two standard multi-view test sequences, Breakdancers and Ballet. Both sequences contain high-motion objects with static backgrounds. Four cameras were used for both sequences at CIF spatial resolution and a temporal resolution of 15 fps. An RTCP turbo encoder with two identical 1/2 rate constituent convolutional encoders, a generator matrix of $\left[1 \quad \left(1+D+D^3+D^4\right)/\left(1+D^3+D^4\right)\right]$ and a random puncturing pattern were used. The results presented herein were produced using a puncturing period of 32. The maximum number of iterations of the turbo decoder was set to 18 and the BER threshold to $10^{-3}$. The KEY H.264 QP values of 40, 36, 32 and 28 were paired with 2, 4, 8, and 16 levels of pixel domain WZ quantization, after heuristic studies that showed that these pairs led to reconstructed data of similar quality. For transform-domain coding, the KEY QP values are matched with the first four quantization matrices of [10]. Results shown are for the whole coded sequence (i.e. including both KEY and WZ data). Comparison are made with H.264 Intra and H.264 predictive coding with zero motion and a GOP of 2 (IBI no Motion). The interleaving block size was 16x16.

### A. Results

The rate-distortion performance of the proposed MV SPI-DVC without any of the enhancement techniques is shown in Fig. 12. Pixel-domain (PD) MV SPI-DVC performs worse than H.264 Intra by 3 and 1.5 dB at high bitrates for Breakdancers and Ballet respectively. Transform-domain coding (TD) MV SPI-DVC performs better with a 0.5 and 1.5 dB improvement in PSNR compared to H.264 Intra for Breakdancers and Ballet respectively (or equivalently a bitrate reduction of up to 20% and 25%). Compared to H.264 Inter no Motion, there is still a gap in performance of up to 1 dB and 2 dB for the Breakdancers and Ballet sequences respectively.

The effect of the GOP size on the performance of the TD MV SPI-DVC is investigated in Fig. 13 for GOPs of length 3 and 5 using the minimized delay structure. The results for Breakdancers show insignificant differences at low bitrates and a slight performance deterioration (0.25dB) at high bitrates. In contrast a GOP of 5 benefits the performance for the Ballet sequence by the same amount at all bitrates. The Breakdancers sequence contains fast and unusual motion causing difficulties to the SI generation process, which thus benefits from close spacing of the KEY frames.

When selective feedback is employed the performance of the proposed codec is improved slightly in transform-domain coding (Fig. 14), and significantly in pixel-domain coding (not shown). Selective feedback requires separate encoding of groups of WZ blocks which together with the coefficient band grouping in TD DVC, result in very small frame lengths fed to the turbo encoder. This affects the performance negatively, as does the need for a feedback map in the case of the simple encoder. In pixel domain coding a 5% and 10% bitrate reduction was recorded for Breakdancers with the simple and smart encoder respectively (3% and 7% for Ballet).

Frame subtraction clearly improves the performance of the codec further (Fig. 14 - 8% and 25% bitrate reduction for Breakdancers and Ballet respectively) closing the gap with H.264 Inter no motion, at the cost of increased encoder complexity. The bitplane projection technique shows slight improvement at low bitrates, but not at higher rates (Fig. 15). If the parity bits of the first bitplane (BP proj+1BP) are additionally requested this performance loss at higher rates is mitigated.

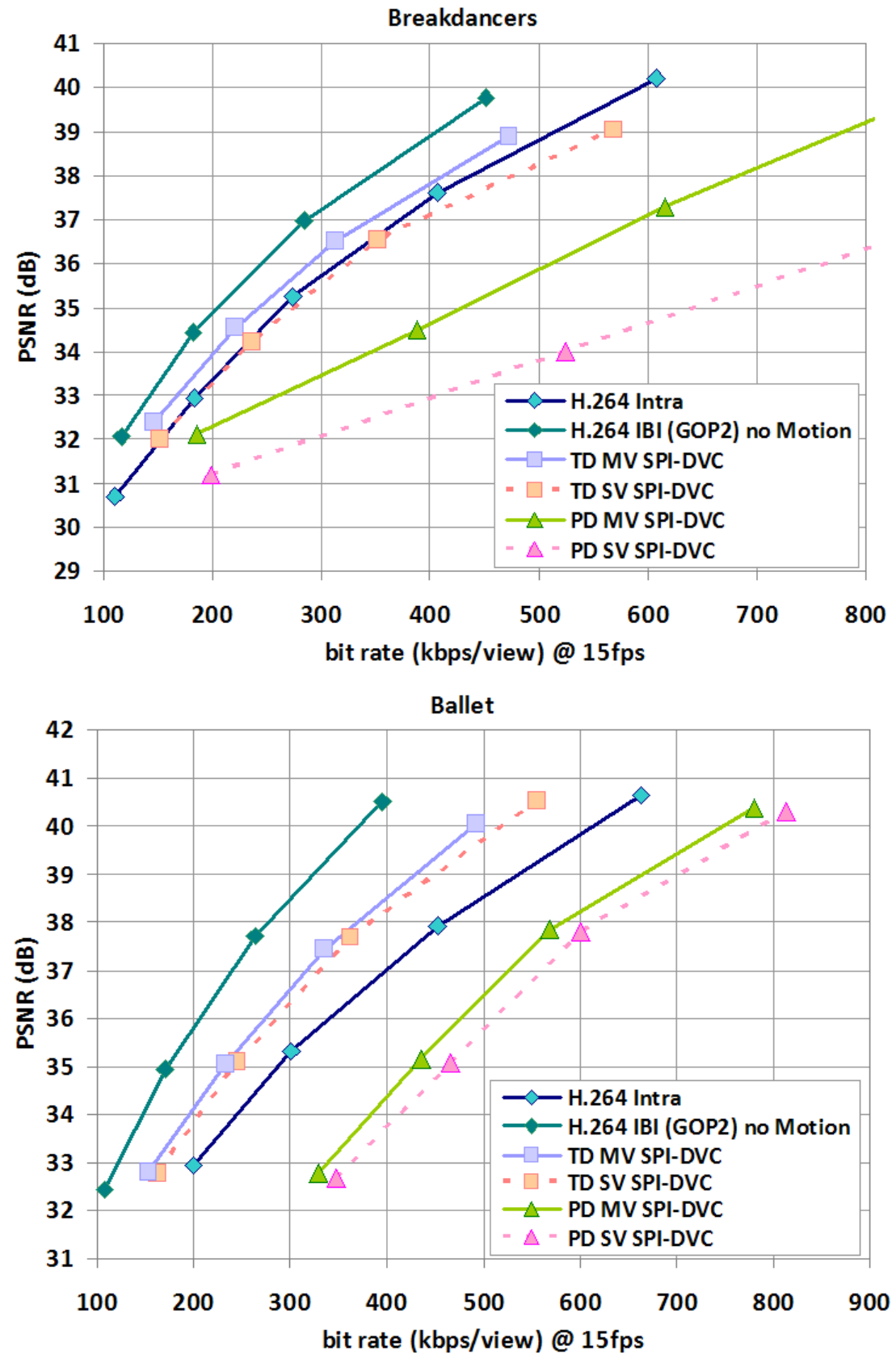

**Fig. 12 Rate distortion performance of PD and TD MV SPI-DVC. Comparison with H.264 Intra and Inter with zero motion.**

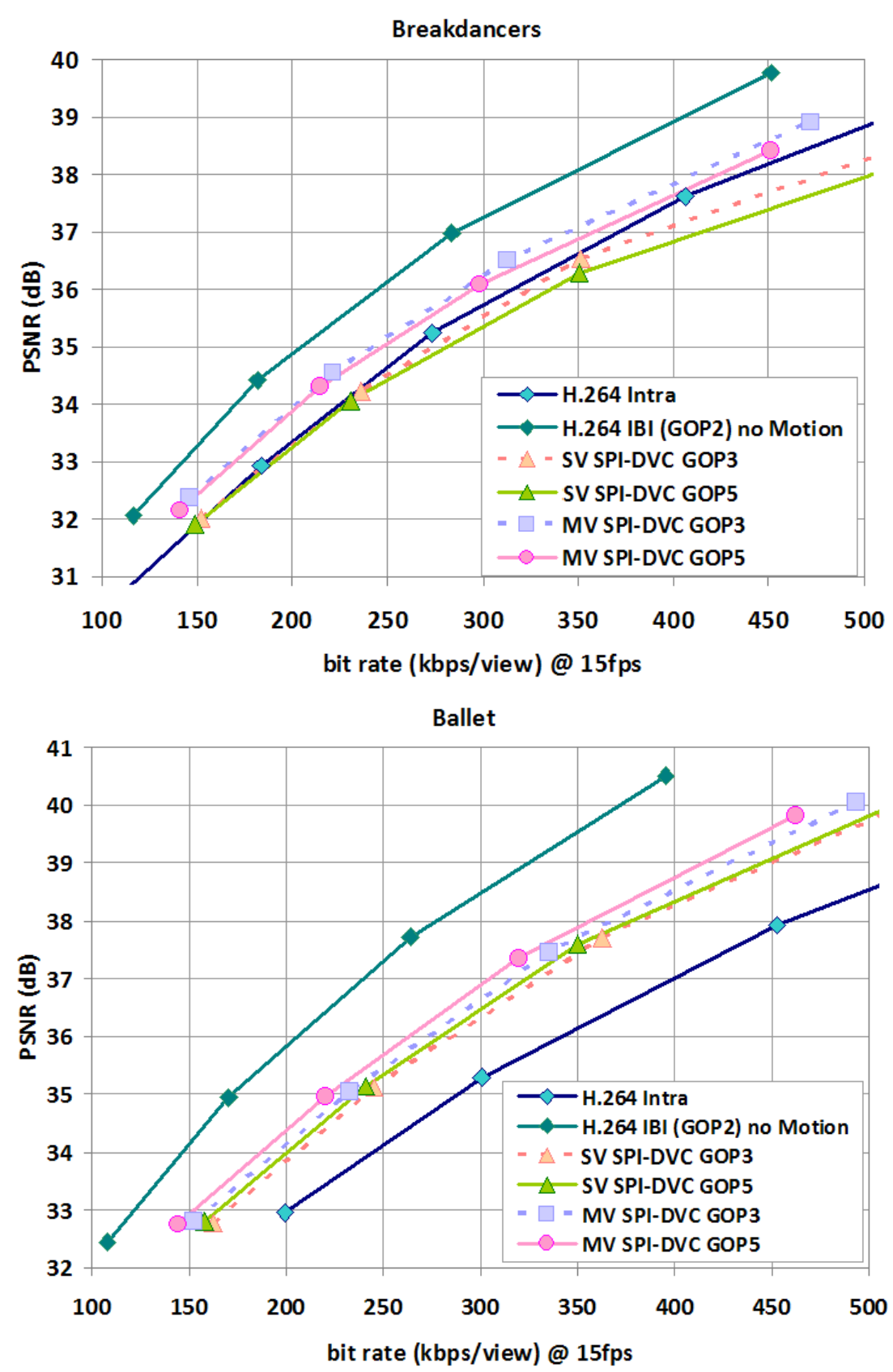

**Fig. 13 Effect of GOP size on the performance of the TD MV SPI-DVC.**

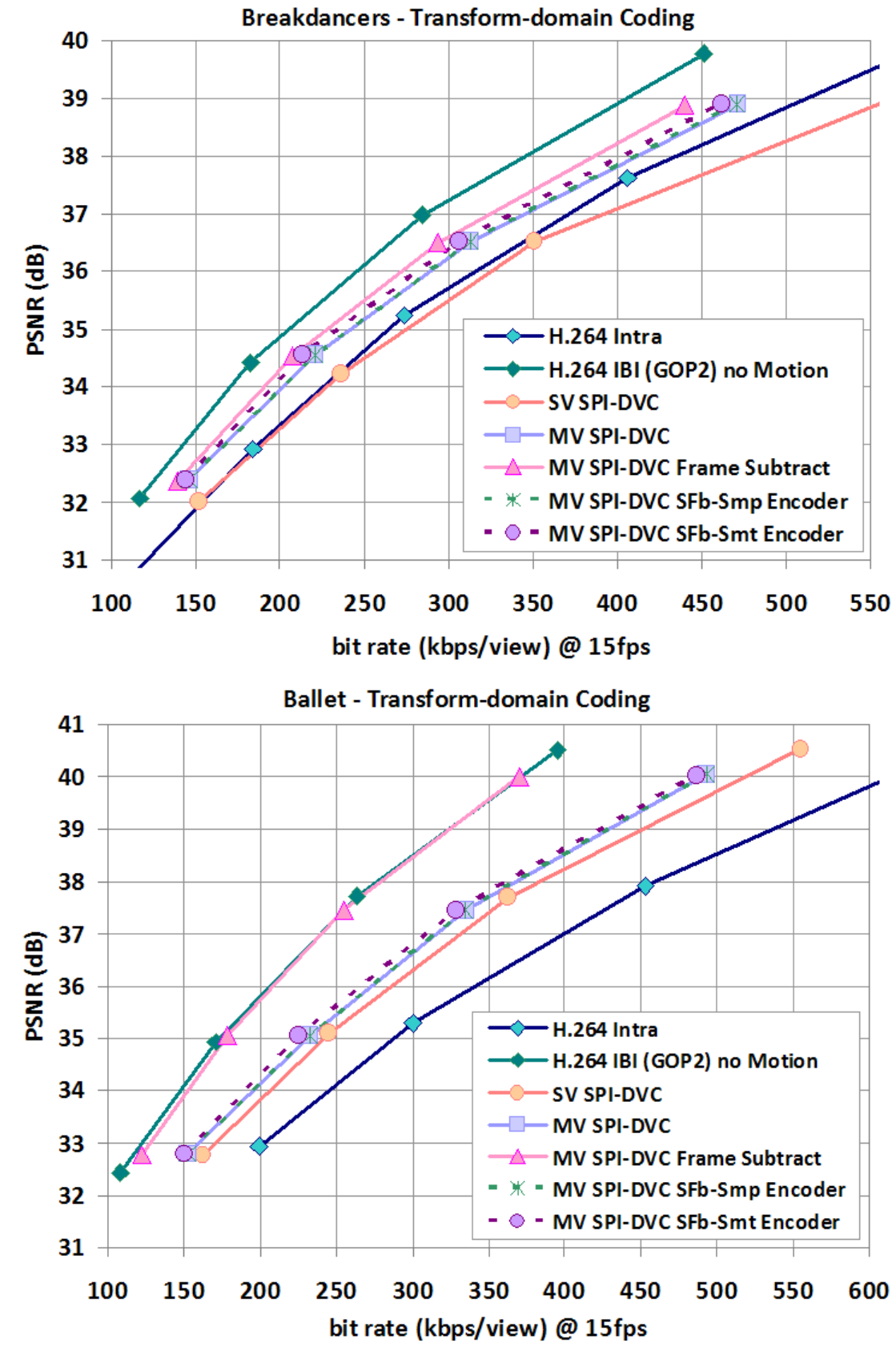

**Fig. 14 Rate distortion performance of TD MV SPI-DVC with selective feedback (SFb) and frame subtraction.**

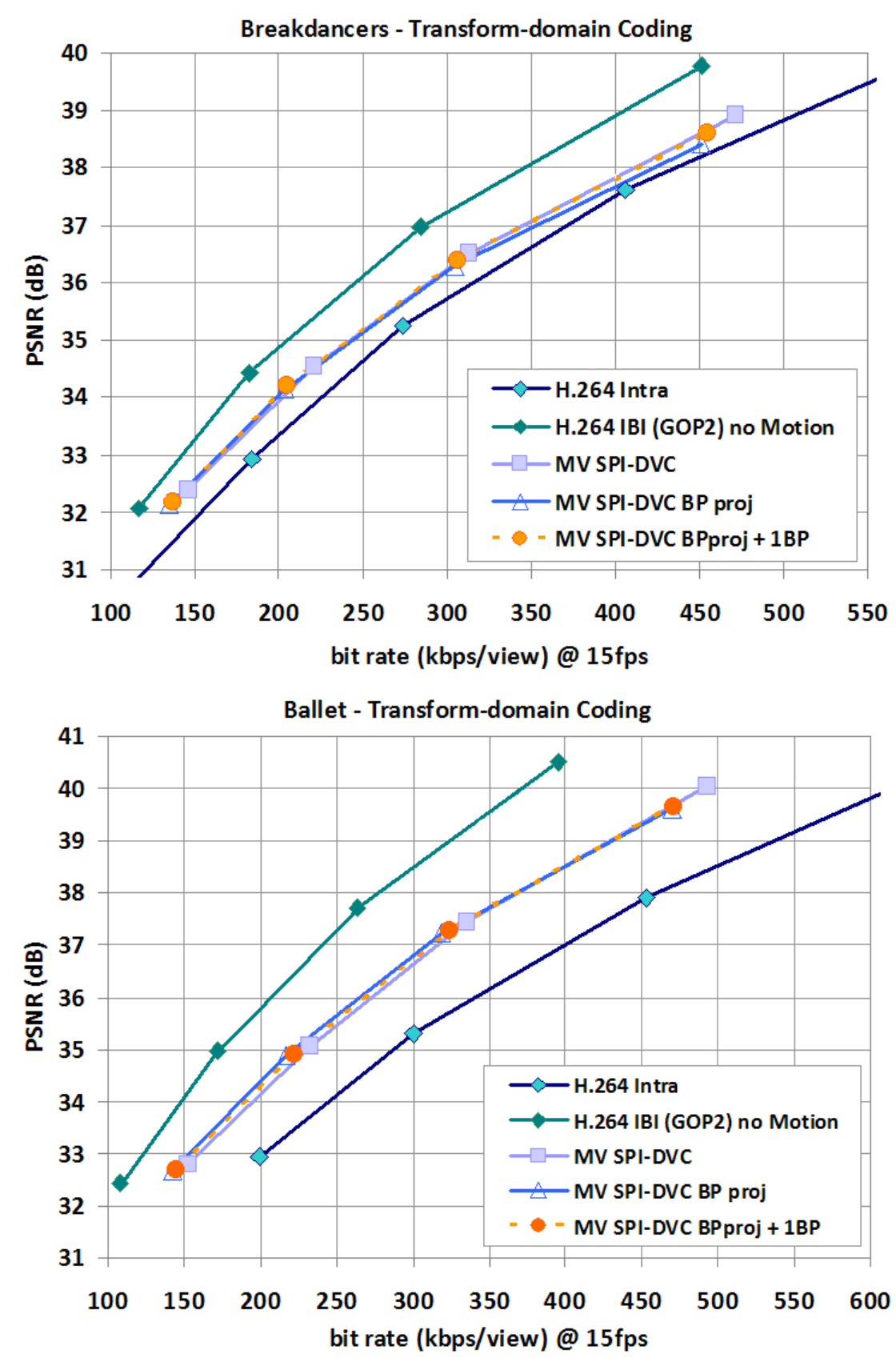

**Fig. 15 Rate distortion performance of TD MV SPI-DVC with bitplane(BP) projection.**

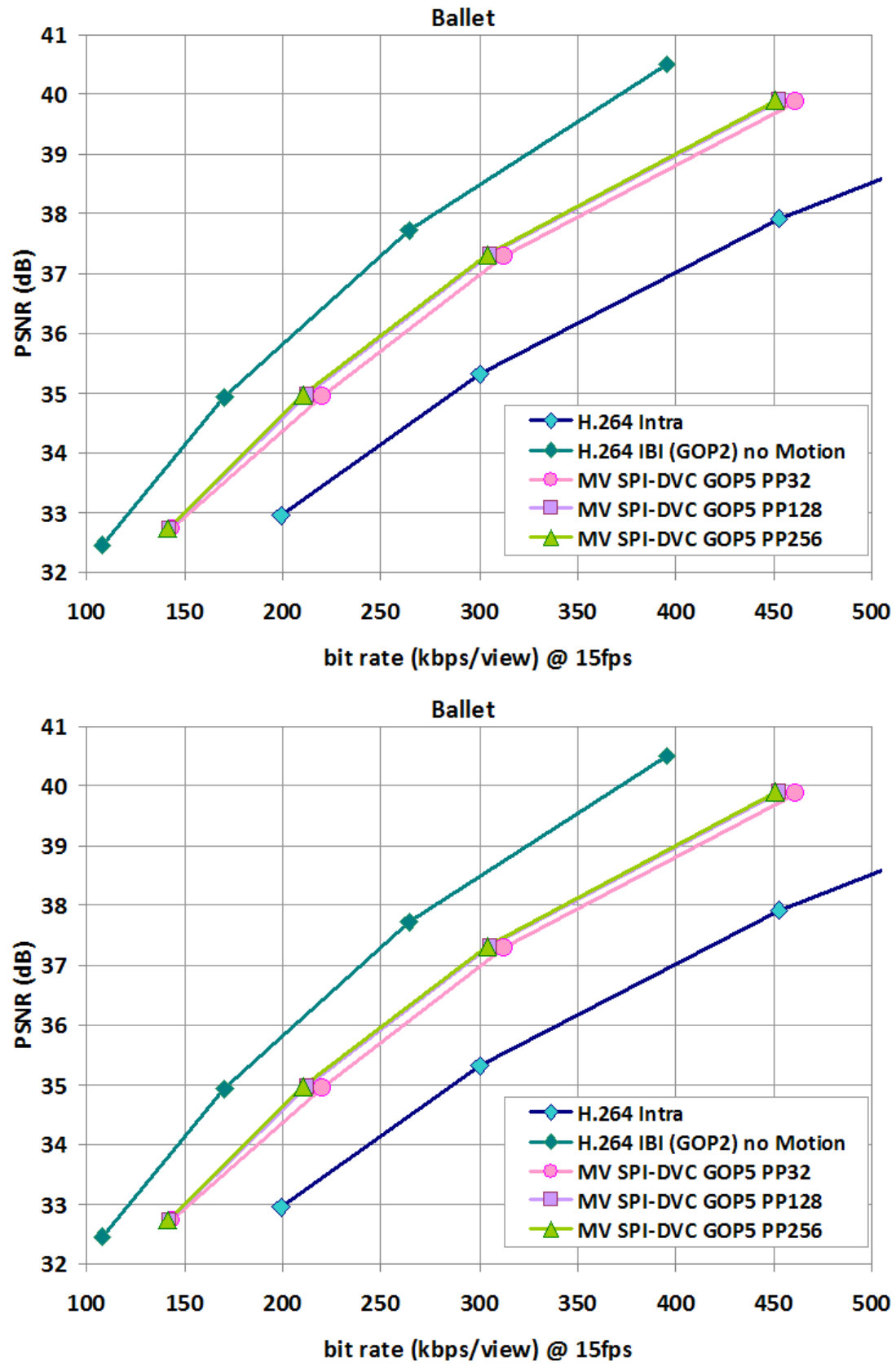

**Fig. 16 Effect of puncturing period on the performance of the TD MV SPI-DVC.**

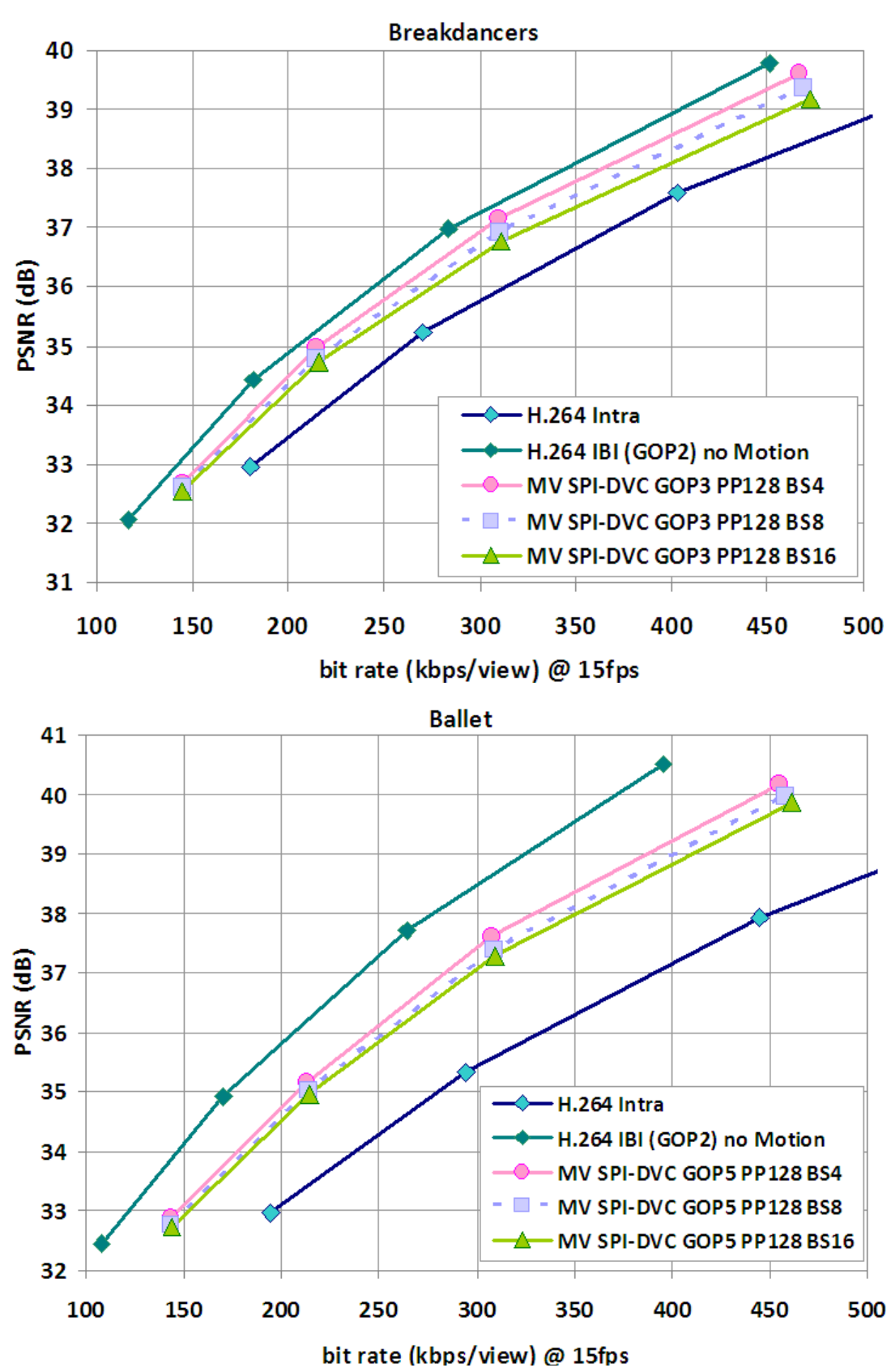

**Fig. 17 Effect of interleaving block size on the performance of the TD MV SPI-DVC.**

A larger puncturing period and a smaller interleaving block size can benefit the performance as shown in Fig. 16 and Fig. 17. With a puncturing period of 128 and an interleaving block size of 4x4, TD MV SPI-DVC closes the gap with H.264 IBI (~0.25dB difference for Breakdancers) and performs significantly better than H.264 Intra (~2dB for Ballet).

### B. *Discussion*

We have presented a novel distributed multi-view video coding framework with Hybrid KEY/WZ frames employed via spatio-temporal-view interleaving of blocks. Each view is encoded completely independently from the others, while the centralized decoder reconstructs multi-view video sequences based on block-based error concealment. Our approach allows good SI generation performed at the block level. Intra- and inter- view data are exploited to generate the side information. We proposed three enhancement techniques for improving the performance at the cost of a slightly increased system complexity. The first technique prioritizes data in a frame. The second technique detects redundant data between frames and the third one employs bitplanes from neighboring views. Results show significant improvement over H.264 Intra coding and comparable performance with H.264 predictive coding with zero motion when frame subtraction is employed.

The attractiveness of DVC lies in the low complexity of the encoder, which is lower than that of H.264 intra coding given that spatial prediction, mode selection, RD optimization and context adaptive entropy coding are only applied (if used) to almost half of the data, and even more so compared to inter coding that additionally uses motion estimation. This of course comes at the cost of (significantly) increased decoder complexity, given that the decoder has to perform error correction (turbo decoding) in addition to motion estimation (see the work of Belkura and Sikora on the complexity of distributed video coding using Turbo codes [26]). However in a scenario where multiple power-limited remote video sensors need to send information to a base station, low complexity, energy efficient encoding becomes important. In such a scenario we can envisage the use of complexity adjustable encoders that switch between different operational modes sourced from both the DVC and the H.264 domain, based on rate-distortion, complexity and error resilience criteria. In this paper we addressed the rate distortion part by proposing a spatio-temporal-view interleaved framework that allows exploitation of all forms of correlation present in the received sequences and which leads to improved performance. Although we could not compare this performance relative to other multi-view DVC schemes, the results we presented relative to H.264 Intra can be used as a guide for comparisons.